\documentclass{emulateapj}

\newcommand{\citepeg}[1]{\citep[{e.g.,}][]{#1}}
\def\Swift{{\textit{Swift}}\,}

\shorttitle{Ultraluminous, Dusty GRB 080607}
\shortauthors{Perley et al.}

\begin{document}

\title{Monster in the Dark: The Ultraluminous GRB 080607 and its Dusty Environment}

\def\ucb{1}
\def\clemson{2}
\def\michigan{3}
\def\ucsc{4}
\def\tautenburg{5}
\def\steward{6}
\def\ucsb{7}
\def\texas{8}
\def\mail{*}

\author{D.~A.~Perley\altaffilmark{\ucb,\mail},
        A.~N.~Morgan\altaffilmark{\ucb},
        A.~Updike\altaffilmark{\clemson},
        F.~Yuan\altaffilmark{\michigan},
        C.~W.~Akerlof\altaffilmark{\michigan},
        A.~A.~Miller\altaffilmark{\ucb},
        J.~S.~Bloom\altaffilmark{\ucb},
        S.~B.~Cenko\altaffilmark{\ucb},
        W.~Li\altaffilmark{\ucb},
        A.~V.~Filippenko\altaffilmark{\ucb},
        J.~X.~Prochaska\altaffilmark{\ucsc},
        D.~A.~Kann\altaffilmark{\tautenburg},
        N.~R.~Butler\altaffilmark{\ucb},
        P.~Christian\altaffilmark{\ucb},
        D.~H.~Hartmann\altaffilmark{\clemson},
        P.~Milne\altaffilmark{\steward},
        E.~S.~Rykoff\altaffilmark{\ucsb},
        W.~Rujopakarn\altaffilmark{\steward},
        J.~C.~Wheeler\altaffilmark{\texas}, and
        G.~G.~Williams\altaffilmark{\steward}
        }

\altaffiltext{\ucb}{Department of Astronomy, University of California,
  Berkeley, CA 94720-3411, USA.}
\altaffiltext{\clemson}{Department of Physics and Astronomy, Clemson
  University, Clemson, SC 29634-0978, USA.}
\altaffiltext{\michigan}{University of Michigan, Randall Laboratory of
  Physics, 450 Church Street, Ann Arbor, MI 48109-1040, USA.}
\altaffiltext{\ucsc}{Department of Astronomy and Astrophysics,
  UCO/Lick Observatory, University of California, 1156 High Street,
  Santa Cruz, CA 95064, USA.}
\altaffiltext{\tautenburg}{Thuringer Landessternwarte Tautenburg,
  Sternwarte 5, D-07778 Tautenburg, Germany.}
\altaffiltext{\steward}{Steward Observatory, University of Arizona,
  933 North Cherry Avenue, Tucson, AZ 85721, USA.}
\altaffiltext{\ucsb}{Physics Department, University of California at
  Santa Barbara, 2233B Broida Hall, Santa Barbara, CA 93106, USA.}
\altaffiltext{\texas}{Department of Astronomy, University of Texas,
  Austin, TX 78712, USA.}
\altaffiltext{\mail}{e-mail: dperley@astro.berkeley.edu .}

\slugcomment{Submitted to AJ 2010-08-31}

\begin{abstract}

We present early-time optical through infrared photometry of the
bright gamma-ray burst GRB~080607, starting only 6~s following the
initial trigger in the rest frame.  Complemented by our previously
published spectroscopy, this high-quality photometric dataset allows
us to solve for the extinction properties of the redshift 3.036
sightline, giving perhaps the most detailed information on the ultraviolet
continuum absorption properties of any sightline outside our Local
Group to date.  The extinction properties are not adequately modeled
by any ordinary extinction template (including the average Milky Way,
Large Magellanic Cloud, and Small Magellanic Cloud curves), partially
because the 2175~{\AA} feature (while present) is weaker by about a
factor of two than when seen under similar circumstances locally.
However, the spectral energy distribution is exquisitely fitted by the
more general Fitzpatrick \& Massa (1990) parameterization of
Local-Group extinction, putting it in the same family as some peculiar
Milky Way extinction curves.  After correcting for this (considerable,
$A_V = 3.3 \pm 0.4$ mag) extinction, GRB~080607 is revealed to have
been among the most optically luminous events ever observed,
comparable to the naked-eye burst GRB~080319B.  Its early peak time
($t_{\rm rest} < 6$ s) indicates a high initial Lorentz factor
($\Gamma > 600$), while the extreme luminosity may be explained in
part by a large circumburst density.  Only because of its early high
luminosity could the afterglow of GRB~080607 be studied in such detail
in spite of the large attenuation and great distance, making this
burst an excellent prototype for the understanding of other highly
obscured extragalactic objects, and of the class of ``dark'' GRBs in
particular.

\end{abstract}


\section{Introduction}

The most extreme gamma-ray bursts (GRBs) have often been the most
illuminating --- both literally and figuratively.  The enormous
isotropic-equivalent energy of GRB 971214 (redshift $z = 3.43$,
$E_{\rm iso} = 3 \times 10^{53}$ erg;
\citealt{Ramaprakash+1998,Odewahn+1998,Kulkarni+1998}) emphatically
demonstrated the need for collimation to bring the energy budget of
long-duration GRBs within physically reasonable values.  Observations
of the mag 9 optical flash of GRB 990123 ($z = 1.61$, $E_{\rm iso} =
3.4 \times 10^{54}$ erg; \citealt{Akerlof+1999,Kulkarni+1999})
anticipated the utility of GRBs to probe the high-redshift universe:
similar events would be easily detectable even at $z>6$.  This
possibility was first vindicated by GRB~050904 ($z=6.29$, $E_{\rm iso}
= 1.2 \times 10^{54}$ erg; \citealt{Kawai+2006,Sugita+2008}), which
for three years remained the most distant GRB known and, at the time,
was also the most luminous optical transient observed in the Universe
\citep{Kann+2007}.  The latter record has since been surpassed
dramatically by GRB~080319B ($z = 0.937$, $E_{\rm iso} =
1.3\times10^{54}$ erg), whose optical afterglow peaked at $V \approx
5$ mag \citep{Racusin+2008,Bloom+2009,Wozniak+2009}.  The current
record for the bolometric isotropic-equivalent energy is held by the
{\it Fermi} burst GRB~080916C ($E_{\rm iso} = 6.5 \times 10^{54}$ erg;
\citealt{Abdo+2009,Greiner+2009}).

Joining this list of record setters is GRB~080607 ($z = 3.036$;
\citealt{Prochaska+2009}), with $E_{\rm iso} = 1.87 \times 10^{54}$
erg \citep{GCN7862}.  This event is remarkable not only for its
intrinsic properties, but also because of its unusual environment: a
Keck spectrum obtained starting only 20 min after the burst
\citep{Prochaska+2009} reveals that the sightline penetrates a giant
molecular cloud in the host galaxy, obscuring the rest-frame visible
light by $A_V \approx 3$ mag of extinction (or $\sim 6$ mag at
1600~\AA, corresponding to the observed $R$ band) before it even began
its journey through intergalactic space.\footnote{Dust extinction is
  limited or absent for the vast majority of well-studied GRBs
  \citep{Schady+2007,Kann+2010}, and no other GRB displays firm
  evidence for molecular lines.} In spite of this extreme attenuation,
the event was bright enough to be detected by small optical telescopes
for over an hour.

The spectroscopic properties of this event have been previously
discussed by \cite{Prochaska+2009}, along with a preliminary analysis
of its extinction properties; further analysis of the spectra was also
presented by \cite{Sheffer+2009}.  In this paper, we analyze several
other aspects of this burst, from the prompt emission (and
simultaneous optical detection) through a late-time search for the
host galaxy, and we present a significantly expanded discussion of its
extinction properties.  In \S \ref{sec:obs} we describe our early-time
multicolor observations of the afterglow with several different
robotic telescopes.  We analyze the optical light curve in \S
\ref{sec:lc}--\ref{sec:earlylc} and show no correlation between the
prompt emission behavior and the early optical observations, starting
at only 6~s post-trigger in the host frame, and we present limits on
color variations at early times.  In \S
\ref{sec:sed}--\ref{sec:extbeta} we examine in more detail the
combined photometric and spectroscopic spectral energy distribution
(SED) and place our final constraints on the host-galaxy extinction
properties, demonstrating the firm detection of a 2175~\AA\ bump, the
highest-redshift detection of this signature to date.  The X-ray light
curve is analyzed in \S \ref{sec:xscat} to search for evidence of dust
scattering in the host at these wavelengths.  In \S \ref{sec:lum} we
place GRB~080607 and its environment in the context of other GRBs,
both ultraluminous and bright events like GRB~080319B as well as the
poorly understood class of extremely dark bursts
\citep{Jakobsson+2004}.  Finally, in \S \ref{sec:physics} we attempt
to explain the origin of the burst's extreme luminosity.

\section{Observations}
\label{sec:obs}

\subsection{Swift}

\begin{figure}
\centerline{
\includegraphics[scale=0.7,angle=0]{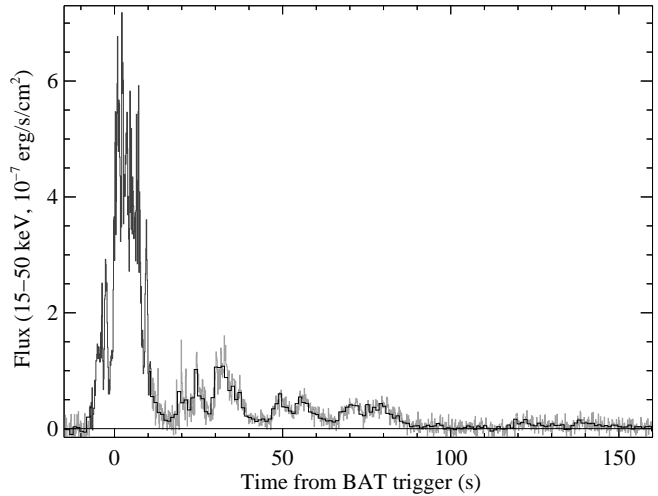}}
\caption{Gamma-ray light curve of GRB~080607 (at a combination of
  128~ms and 1~s binning), showing the bright initial pulse complex
  followed by an additional series of pulses lasting for the next
  several minutes.  Original data for the light curve are taken from
  the {\it Swift} Burst Analyser \citep{Evans+2010}.}
\label{fig:batlc}
\end{figure}

GRB~080607 triggered the Burst Alert Telescope (BAT,
\citealt{Barthelmy+2005}) on the {\it Swift} satellite
\citep{Gehrels+2004} at 06:07:27 on 2010 June 6 (UT dates are used
throughout this paper; times are referenced to this trigger time,
although it is important to note that there was significant emission
before this trigger).  The light curve (Figure \ref{fig:batlc}) is
spiky and erratic, exhibiting a dominant peak at $\sim 4$~s as well as
numerous other, fainter peaks ranging from a few seconds before the
trigger out to $\sim$130~s after, when the signal falls below the
background level.  {\it Swift} slewed immediately to the source and
began pointed observations with the X-ray Telescope (XRT,
\citealt{Burrows+2005}) at 82~s, followed by observations with the
Ultraviolet Optical Telescope (UVOT; \citealt{Roming+2005}) beginning
at 100~s.  Observations continued until 1049~s, after which 
{\it Swift} slewed away temporarily, returning to the field at 4226~s.
From then, observations continued intermittently over the next four
days, after which the X-ray flux was too faint for {\it Swift} to 
detect.

The BAT observations were processed using the {\it Swift} HEAsoft 6.5
software package via the burst pipeline script, {\tt batgrbproduct}.
We calculated spectral parameters both directly and using the Bayesian
formalism described by \cite{Butler+2007}.  Fit to this burst alone, a
GRBM \citep{Band+1993} model provides no significant improvement over
a basic power-law fit over BAT's 15--350 keV energy range (photon
index $\Gamma = 1.16$), suggesting a peak energy above the BAT range.
Using the Bayesian estimate of $E_{\rm peak,obs} = 902_{-460}^{+1170}$
keV and the measured redshift 3.036 \citep{Prochaska+2009}, we
estimate a broad-band isotropic-equivalent energy of $E_{\rm iso} = 2.8^{+1.3}_{-0.9}
\times 10^{54}$ erg.  These values place GRB~080607 second in $E_{\rm iso}$
rank among all {\it Swift} GRBs to date and in the same regime as
extreme events as GRBs 080319B and 990123.

GRB~080607 was observed by other satellites as well (Konus-Wind and
Super-AGILE), enabling a precise measurement of the spectral
parameters.  An in-depth analysis of the Konus data will be presented
in future work by Sbarufatti et al., but preliminary calculations from
\citealt{GCN7862} give the following values: $E_{\rm peak,obs} =
394^{+58}_{-54}$ keV and $E_{\rm iso} = 1.87_{-0.10}^{+0.11} \times
10^{54}$ erg.  These are at the low end of, but generally consistent
with, the Bayesian {\it Swift} result, and confirm that GRB~080607 was
among the most luminous and intrinsically hardest (highest $E_{\rm
  peak,rest}$) GRBs observed by any satellite.
 
The X-ray afterglow was detected throughout the observations; XRT data
were reduced by the procedures of \citet{ButlerKocevski2007}.  The
UVOT afterglow, by contrast, is only marginally detected in the
earliest epoch, and only in White and $V$ filters \citep{GCN7858}.
Both filters are heavily impacted by damped Lyman-$\alpha$ absorption
at $z = 3.036$, and so are not used in our analysis.

\subsection{ROTSE}

The ROTSE-III (Robotic Optical Transient Search Experiment) array is a
worldwide network of 0.45~m robotic, automated telescopes, built for
fast responses to GRB triggers \citep{Akerlof+2003}.
ROTSE-IIIb, located at the McDonald Observatory, Texas, responded
immediately to the initial Gamma-ray Burst Coordinate Network (GCN,
\citealt{Barthelmy+1995}) alert.  The first image started at
06:07:49.0 UT, 22.0~s after the burst,
clearly detecting a bright afterglow at the XRT position in this
exposure.  All ROTSE-III images were processed with our custom RPHOT
photometry program based on the DAOPHOT \citep{stetson87} PSF-fitting
photometry package \citep{qryaa06}. The unfiltered thinned ROTSE-III
CCD has a peak sensitivity in the wavelength range of R band. The
ROTSE magnitudes were thus adjusted using the median offset from the
USNO B1.0 R band measurements of selected field stars.  Observations
are presented (along with photometry from all other telescopes, below)
in Table \ref{tab:phot}.

\subsection{Super-LOTIS}

Super-LOTIS (Livermore Optical Transient Imaging System) is a robotic
0.6~m telescope dedicated to the search for optical counterparts of
GRBs \citep{Williams+2004,Williams+2008}. The telescope is housed in a
roll-off-roof facility at the Steward Observatory Kitt Peak site near
Tucson, AZ.  Super-LOTIS triggered on GRB~080607 and began
observations at 06:08:03 (35~s after the trigger), acquiring a series
of frames in the $R$ band.  The images were reduced and photometry
performed using standard techniques, calibrated relative to nearby
Sloan Digital Sky Survey (SDSS) standard stars.


\begin{deluxetable*}{llccll}
\tabletypesize{\small}
\tablecaption{Photometry of GRB\,080607\label{tab:phot}}
\tablecolumns{6}
\tablehead{
\colhead{Telescope/GCN} &
\colhead{$t$\tablenotemark{a}} & \colhead{Filter} &
\colhead{Exp.~time} &
\colhead{Mag.\tablenotemark{b}} & \colhead{Flux\tablenotemark{c}} \\
\colhead{} &
\colhead{sec} & \colhead{} &
\colhead{sec} & \colhead{} &
\colhead{$\mu$Jy}}
\startdata
PAIRITEL &    89.0 & $J$     &   23.4 & $13.766 \pm 0.107$ & $  5048.6\pm   475.4$ \\
PAIRITEL &    89.0 & $H$     &   23.4 & $12.050 \pm 0.109$ & $ 15657.3\pm  1501.3$ \\
PAIRITEL &    89.0 & $K_s$   &   23.4 & $10.750 \pm 0.139$ & $ 33681.7\pm  4058.1$ \\
KAIT &   188.0 & {\rm clear} &   20.0 & $17.501 \pm 0.055$ & $   363.9\pm    18.0$ \\
KAIT &   158.0 & $I$     &   20.0 & $16.582 \pm 0.094$ & $   635.8\pm    52.7$ \\
KAIT &   128.0 & $V$     &   20.0 & $17.538 \pm 0.142$ & $   437.9\pm    53.7$ \\
ROTSE &    24.5 & {\rm clear} &    5.0 & $14.920 \pm 0.040$ & $  3920.9\pm   141.8$ \\
SuperLOTIS &    40.8 & $R$     &      -- & $15.060 \pm 0.059$ & $  3446.5\pm   182.8$ \\
P60 &   234.8 & $R$     &   60.0 & $17.524 \pm 0.014$ & $   356.3\pm     4.6$ \\
P60 &   406.1 & $i$     &   60.0 & $18.359 \pm 0.022$ & $   186.4\pm     3.7$ \\
P60 &   491.8 & $z$     &   60.0 & $18.694 \pm 0.092$ & $   135.6\pm    11.0$ \\
\enddata 
\tablenotetext{a}{Exposure mid-time, measured from the \emph{Swift} 
trigger (UT 04:08:54).}
\tablenotetext{b}{Observed value, not corrected for Galactic extinction.}
\tablenotetext{c}{Corrected for Galactic extinction ($E_{B-V} = 0.07$ mag).}
\tablecomments{Contains only the first data point in each filter taken by 
each telescope.  A complete table of photometry will be published online.}
\end{deluxetable*}

\subsection{KAIT}

The Katzman Automatic Imaging Telescope (KAIT) at Lick Observatory
\citep{Li+2003} also responded automatically to the {\it Swift} alert
and began taking observations, the first starting at 06:09:25, 118~s
after the BAT trigger.  The KAIT filter sequence consists of a series
of unfiltered observations, followed by a cycle through $V$, $I$, and
unfiltered exposures.  The optical afterglow was detected in all
filters, although it is quite faint in the $V$ band.  Following this
sequence, a series of unfiltered and $I$-band exposures was manually
added, although the afterglow was not detected in the $I$ band and
only marginally detected in our unfiltered exposures at that time
(even after stacking).

Images were reduced using standard techniques.  This left a small
amount of residual on the background sky, which was removed by
subtraction of an illumination frame.  We used aperture photometry to
measure the afterglow flux, calibrating relative to SDSS stars in the
field transformed to the Johnson/Cousins system using the equations of
Lupton (2006).  The clear-band exposures were calibrated to the $R$
band \citep{Li+2003}.

\subsection{PAIRITEL}

The robotic Peters Automatic Infrared Imaging Telescope (PAIRITEL;
\citealt{Bloom+2006}) consists of the 1.3~m Peters Telescope at
Mt. Hopkins, AZ --- formerly used for the Two Micron All Sky Survey
(2MASS; \citealt{Skrutskie+2006}) --- refurbished with the southern
2MASS camera. PAIRITEL uses two dichroics to image in the infrared
(IR) $J$, $H$, and $K_s$ filters simultaneously every 7.8~s.

PAIRITEL responded to the initial BAT alert and slewed immediately to
the source. Observations began at 06:08:44, 77~s after the trigger,
and continued for the next 1.3 hr until the source reached its
hour-angle limit.

The early-time ($<$0.3 hr) raw data files were processed using
standard IR reduction methods via PAIRITEL Pipeline III (Klein et al.,
in prep) and resampled using SWarp \citep{Bertin+2002} to create final
1.0\arcsec\ pixel$^{-1}$ images for final photometry.  Due to changing
sky conditions that complicated the otherwise superior Pipeline III
reductions in the $K_s$ band as the source approached the horizon, the remainder of the
raw data were reduced using an older pipeline which utilized a ``dark
bank'' which more robustly handles flat-fielding in such cases.

PAIRITEL's standard observing cycle is to take three 7.8~s exposures
in immediate succession at each dither position. While the early
afterglow is detected in even the shortest 7.8~s frames, for
signal-to-noise ratio (S/N) and calibration considerations, we report
23.4-s ``triplestacks'' (the median of all three images at each dither
position) as our shortest exposures. These images were further binned
at successively later times to further improve the S/N.

Aperture photometry was performed using custom Python software,
utilizing Source Extractor (SExtractor; \citealt{Bertin+1996}) as a
back end.  Four calibration stars present in all images were chosen
based on brightness, proximity of nearby contaminating sources, and
location relative to bad pixels.  The optimal aperture of
$\sim$3\arcsec\ radius was determined by minimizing the absolute error
relative to 2MASS magnitudes of our four calibration stars.

Calibration was performed by redetermining the zero-point for each
image individually by comparison to 2MASS magnitudes using these four
stars.  The resulting statistical uncertainty in the zero-point
is negligible relative to other sources of error. Additional,
systematic sources of error are addressed in detail by
\cite{Perley+2010}; we use a similar procedure here to determine the
total uncertainty of each point.

\subsection{P60}

The robotic Palomar 60\,inch telescope (P60; \citealt{Cenko+2006})
automatically responded to the \textit{Swift} trigger for GRB\,080607,
executing a predefined sequence of observations in the Kron $R$ and
Sloan $i^{\prime}$ and $z^{\prime}$ filters beginning 174~s after the
burst trigger time.  Individual images were reduced in real time using
standard IRAF\footnote{IRAF is distributed by the National Optical
  Astronomy Observatory, which is operated by the Association for
  Research in Astronomy, Inc., under cooperative agreement with the
  National Science Foundation (NSF).} routines.  The images were
calibrated with respect to several dozen field stars from the SDSS
Data Release 7 \citep{DR7}, using the filter transformations of
\cite{Jordi+2006} for the Kron $R$ filter.  

\subsection{Keck Spectroscopy}

We initiated spectroscopic observations of the afterglow with the
Low Resolution Imaging Spectrometer \citep[LRIS;][]{Oke+1995} on
the Keck I 10~m telescope at
13 min after the {\it Swift} trigger, although due to poor guiding
this first frame was not usable.  The first exposure
used in our analysis began at 20.1 min following the trigger.  
Several additional exposures were taken over
over the next two hours using the B600 grism and both the R400 and
R1200 gratings; our final observations span a wavelength range of
3000--9000~\AA.  Observations were flux-calibrated relative to the
spectroscopic standard HZ~44.  More details on these spectroscopic
observations and our reductions are given by \cite{Prochaska+2009}.

\subsection{Keck Host-Galaxy Imaging}

The field of GRB\ 080607 was imaged in several deep integrations at
Keck in various optical/IR filters from $g$ through $K_s$.  None of
these integrations resulted in a secure detection of the host galaxy,
with the possible exception of the $g$ band, in which a faint source
is detected at about 3$\sigma$ above background at the afterglow
position ($g = 27.4 \pm 0.3$ mag).  A log of our ground-based host
observations is reported in Table \ref{tab:host}.

\begin{deluxetable}{lllll}
\tablewidth{0pc}
\tablecaption{Host-Galaxy Limits}
\tablehead{ \colhead{Instrument} & \colhead{Obs. date} & \colhead{Exp. time} & \colhead{Filter} &  \colhead{5$\sigma$ limit} \\
            \colhead{}           & \colhead{(UT)}      & \colhead{(s)}       & \colhead{} &        \colhead{(mag)}}
\startdata
  Keck I / LRIS    & 2009-02-19  & 2490  & $g$   &  $> 27.3$  \\
  Keck I / LRIS    & 2009-02-19  & 2220  & $I$   &  $> 25.3$  \\
  Keck I / NIRC    & 2009-05-31  & 3600  & $K_s$ &  $> 21.6$  \\
\enddata
\tablecomments{$5\sigma$ limiting magnitudes on a host galaxy at the
  afterglow position from our ground-based optical and IR observations
  at the Keck Observatory.}
\label{tab:host}
\end{deluxetable}

The host galaxy is, however, well detected at 1.6~$\mu$m in a deep
{\it Hubble Space Telescope (HST)} image using WF3, as well as in both
of the warm {\it Spitzer} IRAC channels (3.6~$\mu$m and 4.5~$\mu$m).  The
extreme optical faintness of this system, while partially due simply
to its high redshift ($z=3.036$), makes this galaxy of particular
interest: determination of the redshift would be exceptionally
difficult using traditional field-survey techniques, illustrating
the unique ability of GRBs
to select and study optically faint galaxies at high redshift.
Further discussion of the host galaxy, including detailed analysis of
both the ground- and space-based imaging, will be presented in
upcoming work by Chen et al.

\section{Analysis}

\subsection{Optical Light Curve}
\label{sec:lc}

\begin{figure*}
\centerline{
\includegraphics[scale=0.8,angle=0]{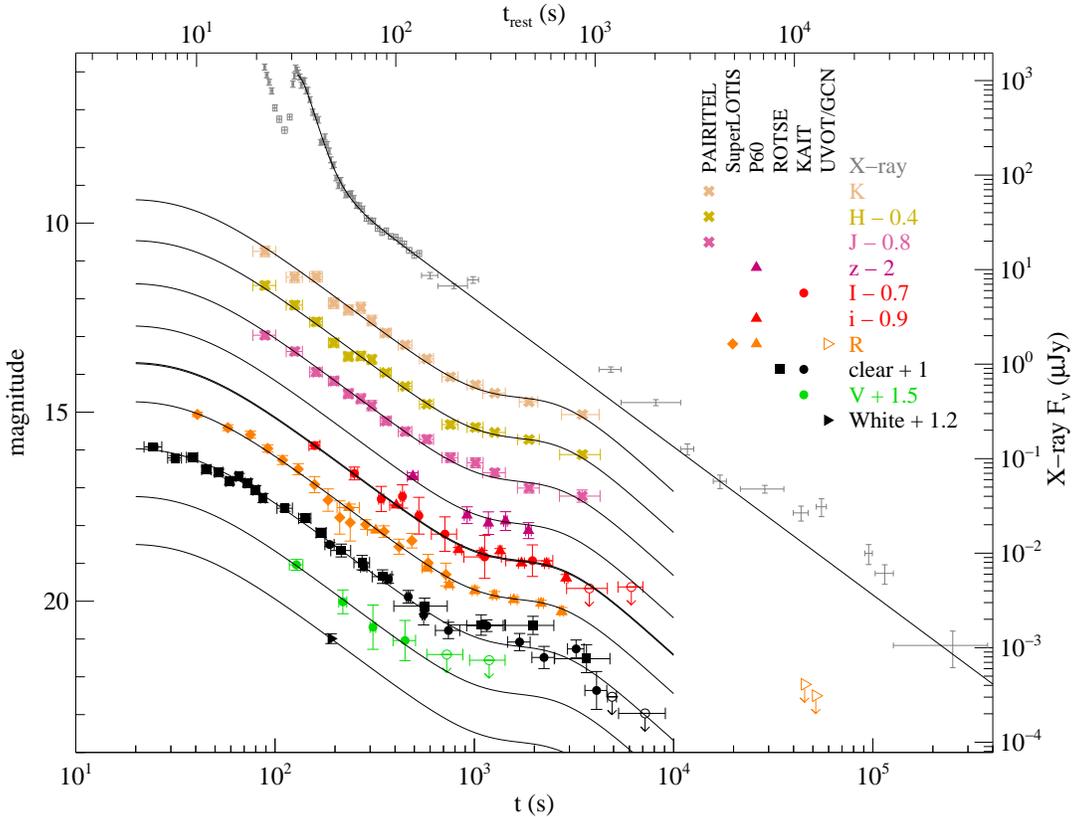}}
\caption{Multi-band light curve of GRB~080607 from a variety of
  ground-based telescopes as well as the \Swift XRT, fit to a sum of
  two broken power laws.  (XRT data are fit to a sum of two unbroken
  power laws.)    Magnitudes are in the Vega system and (with
  the exception of $R$ and $K_s$) have been shifted as indicated for
  clarity; these magnitudes are \emph{not} corrected for Galactic
  extinction (which is nearly insignificant) or host extinction (which
  is very large).  The afterglow initially fades slowly, then
  steepens; it briefly levels out at $10^3$~s before breaking again
  and is not detected after 5000~s.  The late-time $R$-band limits are
  from \cite{GCN7891}.  We use the BAT trigger time for $t_0$, which
  corresponds to the start of the largest prompt-emission pulse; using
  the start of gamma-ray emission instead does not significantly 
  change the qualitative results.}
\label{fig:lc}
\end{figure*}

The multi-band light curve of GRB~080607 is plotted in Figure
\ref{fig:lc}.  After an initially slow decay the light curve steepens
(decay index $\alpha = 1.65$, using the convention $F \propto
t^{-\alpha}$) before flattening out at 1000~s to a temporarily flat
decay.  This slow decay lasts for approximately another hour before
fading again, becoming undetected in our final KAIT exposures.

The light curve was fitted using the techniques described by
\cite{Perley+2010} and previous works by our group, modeling the light
curve as the sum of several broken power laws.  Our temporal coverage
of this event is limited (ending at $10^4$~s), making the analysis
simple: we employ two \cite{Beuermann+1999} broken power laws, one to
describe the early behavior and a second to describe the later
flattening.  Because we do not detect the rising phase of the
afterglow, the pre-break index of the first power-law component is not usefully
constrained by our data and is fixed arbitrarily to $-0.5$; the
falling phase of the second (late) component is similarly not well
constrained and we fix the post-break decay index to $2.0$, the shallowest
value required for consistency with our upper limits in this model.

Modest but significant color change has been previously observed in
early-time GRB afterglows (see \citealt{Perley+2008} or
\citealt{Bloom+2009} for two prominent examples), a possibility which
we model by allowing the intrinsic spectral power-law index $\beta$
($F \propto \nu^{-\beta}$) to vary between components or across
breaks.  However, in the case of GRB~080607, any such color change is
not significant: the change in intrinsic index between the fast-decay
and flat components is only $\Delta\beta = 0.07 \pm 0.07$ and only
modestly improves the goodness of fit.  Therefore, for simplicity we
assume no color change during our observations of this burst.

\subsection{Absence of Optical/High-Energy Correlations}
\label{sec:earlylc}

Our optical follow-up observations of this burst begin extremely
early.  The ROTSE coverage begins at only 21~s after the BAT trigger,
corresponding to less than 6~s in the GRB rest frame.  The prompt
emission was still extremely active at this time: at least five major
gamma-ray flares occurred during our optical observations, the
last of which was also caught at X-ray wavelengths by the XRT.
PAIRITEL, KAIT, and SuperLOTIS were all observing during this last
flare.

\begin{figure}
\centerline{
\includegraphics[scale=0.7,angle=0]{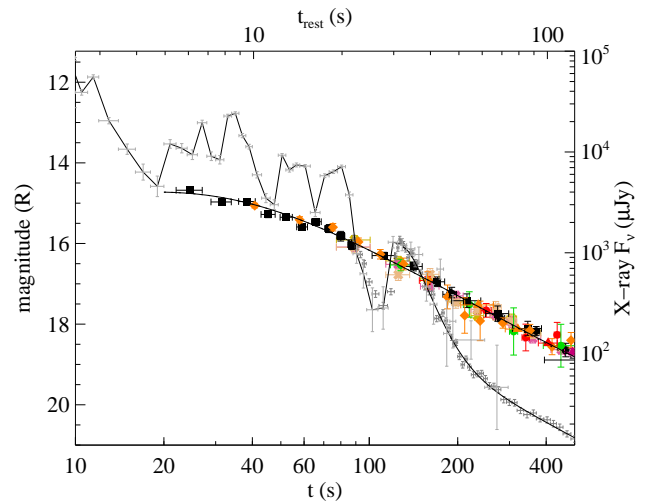}}
\caption{Early-time optical and gamma-ray/X-ray light curve of
  GRB\ 080607, demonstrating the extremely early peak $t_{\rm
    rest} \lesssim 6$~s as well as the absence of any visible
  correlation between the optical and high-energy light curves.
  (Optical fluxes in other bands have been scaled to match the
  $R$ band; the gamma-ray light curve is scaled to match the
  X-ray curve.)  Symbols are the same as in Figure \ref{fig:lc}.}
\label{fig:earlylc}
\end{figure}

Even in this rich overlapping dataset, there is no correlation visible
between the optical and high-energy light curves of the type seen by,
for example, \cite{Vestrand+2005}, \cite{Blake+2005}, and
\cite{Beskin+2010}.  In Figure \ref{fig:earlylc} we overplot the
gamma-ray, X-ray, and optical light curves using the same relative
scaling.  In spite of the erratic, flaring high-energy behavior, we
see no sign of significant deviation of the optical light curves from
their smooth power-law behavior at any point.  This is consistent with
other ROTSE-followed bursts \citepeg{Yost+2007,Rykoff+2009} and
provides another clear example of a burst whose afterglow behavior is
clearly divorced from that of the prompt emission.

The lack of even modest influence of the prompt emission on the
afterglow may initially seem surprising: even if truly prompt
(internal-shock) emission is absent in this band, one might expect
that some of the energy being released so liberally by the central
engine might end up in the external shock, causing a less dramatic but
still observable rebrightening of the afterglow (a refreshed shock;
\citealt{Panaitescu+1998}).  We note, however, that despite the
intense flaring shown in Figure \ref{fig:earlylc}, this emission is
actually dwarfed by an earlier episode: the initial pulse of the
prompt emission (see Figure \ref{fig:batlc}) exceeds any of the later
spikes by an order of magnitude in both intensity and energy, and it
is this initial pulse that dominates the energetics of the burst.  The
later flares are much more modest by comparison, so even presuming
direct input from outflow (revealed by the prompt emission) to
external shock (revealed by the afterglow), the absence of further
brightening is not necessarily surprising.

\subsection{Spectral Energy Distribution}
\label{sec:sed}

Our light-curve fits naturally provide values for the afterglow flux
in each filter at any given time, allowing us to model the SED at any
time during our observations.  Because of the absence of significant
color change, the choice of extraction epoch is arbitrary; 300~s is
chosen in this case (when all ground-based telescopes were observing
and the afterglow was still bright enough to be well detected in all
bands).

The Keck optical spectroscopy covers a wide range of wavelengths and
was carefully flux-calibrated: photometric standard stars were
observed immediately after our observations at similar airmasses and
the night was photometric throughout.  Accordingly, we couple our
spectrum to the photometry to improve the precision of our broadband
modeling.

The optical spectrum is replete with lines from a variety of elements
and molecules at the host-galaxy redshift of 3.036.  The analysis of
these line features is discussed extensively by \cite{Prochaska+2009}
and \cite{Sheffer+2009}, and we will not repeat it here; our primary
interest is in the continuum.  Although the contribution of absorption
lines is usually ignored in GRB photometric dust modeling, the lines
in the spectrum of GRB\ 080607 are so abundant and so strong that
ignoring them would create systematic errors significantly larger than
our photometric uncertainties in both the spectrum itself and in the
broadband photometry.  In addition, nearly the entire spectrum at
wavelengths shorter than $\sim$6900~{\AA} is affected by a forest of
weak lines from vibrationally excited $H_2^*$, further complicating
the analysis.

Fortunately, we are able to correct for these effects.  We use the
line list presented in Table 1 of \cite{Prochaska+2009} to identify
all regions of the spectrum affected by ionic lines, including the
entire spectrum blueward of 5400~{\AA}, which is affected by the host
damped Lyman-$\alpha$ and the Lyman-$\alpha$ forest.  In addition, the
spectrum is corrected for the subtler but more widespread $H_2^*$
absorption using the model developed by \cite{Sheffer+2009}.  We then
fit a sixth-order polynomial to the ionic line-free regions of this
corrected spectrum to create a continuum model and perform synthetic
photometry using both the model spectrum and the observed, uncorrected
spectrum (and take the ratio) to calculate an adjustment factor with
which to convert the observed (line-affected) fluxes to continuum
(line-free) fluxes for each of our broadband filters covering the
optical spectrum ($R$, $I$, $i$, and $V$; we assume the line
contribution is small further to the red).  We also wish to use the
flux-calibrated spectrum itself in later analysis, so we scale the
spectrum to the photometric SED extraction epoch of 300~s (the scale
factor is determined by the value that minimizes $\chi^2$ for our
extinction fits; see \S \ref{sec:ext}) and bin the flux in blocks of
200~{\AA} (excluding line-affected regions).  Uncertainties are
determined by combining the statistical uncertainties from the
spectrum with a systematic term of 3\% per bin to incorporate any
uncertainty in the flux calibration (10\% is used for $<$5500~\AA\ and
$>$9000~\AA, which are especially uncertain.)  Using this technique,
we generate a line-corrected narrowband SED spanning
5400--9200~\AA\ to complement our line-corrected photometry.  The
afterglow fluxes from direct and synthetic photometry are presented in
Tables \ref{tab:sed} and \ref{tab:spec}, respectively.

\begin{deluxetable*}{lllllllll}
\tablewidth{0pc}
\tablecaption{Model Fluxes at $t = 300$ s}
\tablehead{ \colhead{Filter} & \colhead{$\lambda_{\rm obs}$} & \colhead{$F_{\rm obs}$} & \colhead{$m_{\rm obs}$} & \colhead{$A_{\lambda,{\rm Gal}}$} 
            & \colhead{$\Delta m_{\rm lines}$} & \colhead{$m_{\rm cont}$} & \colhead{$F_{\rm cont}$} & \colhead{$A_{\lambda,{\rm host}}$} \\
            \colhead{}       & \colhead{(\AA)}    & \colhead{($\mu$Jy)}    & \colhead{(mag)}    &  \colhead{(mag)} & \colhead{(mag)} & \colhead{(AB mag)}   & \colhead{($\mu$Jy)} & \colhead{(mag)}}
\startdata
X-ray & 12.4  & 34.04            &                &       &       &                   &                  &  \\
$V$   & 5505  & $94.25 \pm 11.4$ & 18.98 $\pm$ 0.13 & 0.07 & 0.60 & 18.29 $\pm$ 0.13   & $174.6 \pm 21.1$ & 6.23 \\
$R$   & 6588  & $200.6 \pm 12.6$ & 17.97 $\pm$ 0.07 & 0.06 & 0.21 & 17.88 $\pm$ 0.07   & $256.6 \pm 16.1$ & 5.76 \\
$i$   & 7706  & $260.5 \pm 16.6$ & 17.86 $\pm$ 0.07 & 0.05 & 0.06 & 17.75 $\pm$ 0.07   & $287.5 \pm 18.3$ & 5.88 \\
$I$   & 8060  & $214.0 \pm 19.1$ & 17.64 $\pm$ 0.10 & 0.04 & 0.07 & 17.97 $\pm$ 0.10   & $236.0 \pm 21.0$ & 6.05 \\
$z$   & 9222  & $242.5 \pm 21.4$ & 17.96 $\pm$ 0.10 & 0.03 & 0    & 17.91 $\pm$ 0.10   & $249.8 \pm 22.0$ & 6.12 \\
$J$   & 12350 & $867.4 \pm 51.4$ & 15.66 $\pm$ 0.06 & 0.02 & 0    & 16.54 $\pm$ 0.06   & $883.1 \pm 52.3$ & 4.94 \\
$H$   & 16620 & $2296 \pm 131$   & 14.12 $\pm$ 0.06 & 0.01 & 0    & 15.48 $\pm$ 0.06   & $2322  \pm 132$  & 4.20 \\
$K_s$ & 21590 & $5866 \pm 337$   & 12.64 $\pm$ 0.06 & 0.01 & 0    & 14.47 $\pm$ 0.06   & $5911  \pm 339$  & 3.30 \\
\enddata
\tablecomments{Broadband afterglow fluxes as determined by the
  light-curve model, interpolated to $t=300$~s after the trigger.
  Observed magnitudes are not corrected for Galactic extinction and
  are in the Vega system (except for the SDSS $i$ and $z$ filters,
  which are given in the SDSS filter system.)  Continuum magnitudes
  and fluxes have been corrected for both Galactic extinction (from
  NED) and line absorption (calculated using our optical Keck
  spectroscopy).}
\label{tab:sed}
\end{deluxetable*} 

\begin{deluxetable}{ll}
\tablewidth{0pc}
\tablecaption{Binned, Line-Interpolated Keck Spectroscopy}
\tablehead{ \colhead{$\lambda$} & \colhead{$F_{\nu,{\rm cont}}$} \\
            \colhead{(\AA)}       & \colhead{($\mu$Jy)} }
\startdata
5448.56  &  128.24 $\pm$ 0.94 \\
5670.14  &  146.45 $\pm$ 1.79 \\
5842.98  &  160.54 $\pm$ 2.48 \\
6112.85  &  201.91 $\pm$ 2.52 \\
6235.50  &  196.32 $\pm$ 1.74 \\
6476.67  &  211.72 $\pm$ 2.45 \\
6776.25  &  233.69 $\pm$ 1.48 \\
7099.05  &  240.99 $\pm$ 1.72 \\
7281.72  &  238.65 $\pm$ 1.04 \\
7483.82  &  231.69 $\pm$ 0.97 \\
7774.03  &  213.43 $\pm$ 2.41 \\
7997.30  &  195.15 $\pm$ 0.84 \\
8181.51  &  187.27 $\pm$ 0.85 \\
8411.10  &  180.74 $\pm$ 2.26 \\
8607.66  &  165.18 $\pm$ 1.02 \\
8759.01  &  153.90 $\pm$ 6.60 \\
8892.83  &  151.97 $\pm$ 4.61 \\
9044.91  &  162.99 $\pm$ 2.90 \\
9175.62  &  158.64 $\pm$ 2.36 \\
\enddata
\tablecomments{Uncertainties are photometric only and do not include
  any systematic term.  Fluxes are corrected for $H_2^*$ absorption 
  in the host galaxy and for Galactic extinction.}
\label{tab:spec}
\end{deluxetable} 

\subsubsection{Extinction Fitting}
\label{sec:ext}

\begin{figure}
\centerline{
\includegraphics[scale=0.7,angle=0]{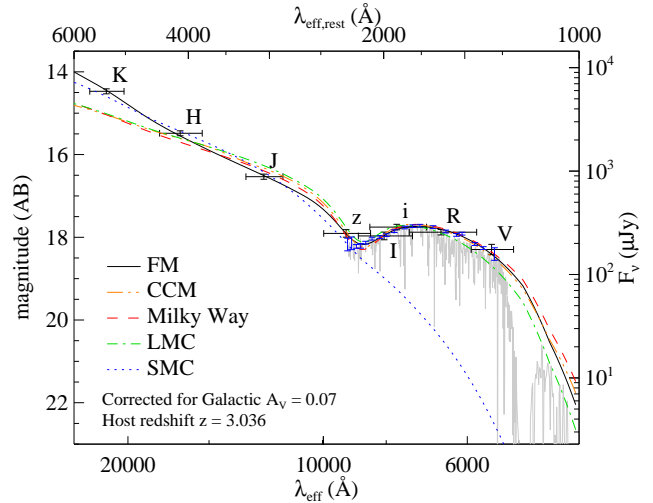}}
\caption{The combined photometric and spectroscopic SED of GRB\ 080607
  fitted with several different extinction models.  (FM = Fitzpatrick
  \& Massa 1990, CCM = Cardelli, Clayton, \& Mathis 1989).  Black
  error bars indicate broadband photometry; blue error bars show the
  binned pseudo-photometry as derived from the Keck spectrum and
  corrected for line absorption (including $H_2^*$).  The light-grey
  line shows the spectrum (mostly unbinned and including all lines).
  Several different extinction fits are shown; only the general FM
  model (solid black) is an acceptable fit to the data.  The SMC curve
  shown is a fit to the IR data only (an SMC fit to all data converges
  to $A_V = 0$ mag).}
\label{fig:sed}
\end{figure}

The combined photometric and spectroscopic SED is plotted in Figure
\ref{fig:sed}.  It is immediately evident that this curve is unlike
almost any other GRB SED that has been observed in detail to date.
First, the color is extremely red: a power-law fit to the broadband
photometry would give a spectral slope ($F \propto \nu^{-\beta}$) of
$\beta \approx 3$, at odds with the theoretically expected value of
$\beta = 0.5$--1.2 for an early fading afterglow \citep{Sari+1998}.
Second, it is not monotonic: the flux drops sharply from the $K_s$ band
until $\sim$2200~{\AA} in the rest frame before actually recovering,
showing a local maximum at $\sim$1600~{\AA} before falling again
further to the blue.

These properties are immediately recognizable as signatures of dust
extinction, and particularly of Milky-Way like extinction with its
broad 2175~{\AA} absorption band.  This strong extinction imprint, in
combination with our high-S/N afterglow observations spanning the
entire optical/near-IR window, permits analysis of the rest-frame UV
extinction properties at a level of detail that is almost never
possible with GRBs (or indeed, with any other technique at this
redshift range).

To constrain the dust properties, we initially followed the standard
procedure for GRB extinction measurements by fitting the average Milky
Way (MW), Large Magellanic Cloud (LMC), and Small Magellanic Cloud
(SMC) curves, assuming an intrinsic power-law sectrum.  (Here, and
elsewhere unless otherwise specified, the intrinsic spectral slope
over the optical range is fixed at $\beta = 0.7$.  Fortunately,
because the amount of extinction for this burst is so large,
deviations from this assumption do not significantly affect our
results, except to slightly increase the uncertainties in the derived
parameters, as we will discuss in \S\ref{sec:extbeta}.)  In all three
cases we use the \cite{Fitzpatrick1999} parameterization of
Local-Group extinction as implemented in the GSFC IDL package, with
$R_V$ fixed to their average value for each galaxy; for SMC extinction
we use the Fitzpatrick parameters from \cite{Gordon+2003} (SMC bar
average).  SMC extinction is ruled out (it converges to $A_V=0$ mag
with $\chi^2$/dof = 1159/24), as it rises steeply to the far-UV (FUV)
and does not allow for the 2175~{\AA} bump feature that is so
prominent in our data.  The LMC and MW curves fit the data much
better, but nevertheless they are not statistically acceptable either.
Both curves are too flat in the observed IR; the MW curve also
significantly overestimates the strength of the 2175~{\AA} bump.

This should not be a surprise: even within our own Galaxy a
significant diversity of extinction laws is evident.  The majority of
observed Galactic sightlines are consistent with variation in a single parameter $R_V$, which
describes the relative ``greyness'' (wavelength independence) of the
extinction at optical through UV wavelengths (\citealt{CCM}; hereafter
CCM).  A small number of sightlines in the MW (and all sightlines
within the LMC and SMC) require additional parameters to fit
accurately.  A more general Local-Group extinction law, developed by
Fitzpatrick \& Massa (1990; hereafter FM), is able to fit essentially
all local sightlines by adding an additional family of parameters:
$c_2$ for further variations in steepness in the UV, $c_3$ for the
strength of the 2175~{\AA} bump, $\gamma$ for the bump's width, and
$c_4$ for the strength of the FUV rise.  (The parameter $c_1$ is also
present in principle, but it is essentially degenerate with $c_2$ and
$R_V$, and in practice it is fixed based on those values.  In
addition, the parameter $x_0$ describes the central wavelength of the
2175~{\AA} bump, but it has not been conclusively shown to vary and is
fixed to the average value.)

\begin{figure}
\centerline{
\includegraphics[scale=0.7,angle=0]{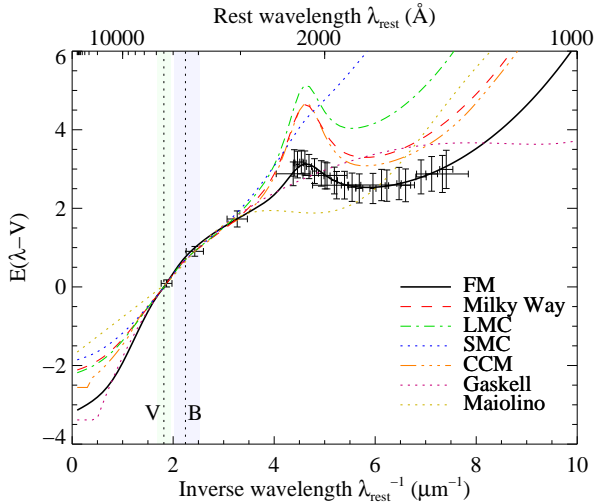}}
\caption{Different extinction curves shown against our afterglow data,
  shown as the selective extinction $E(\lambda-V) = A_{\lambda}-A_{\rm
    V}$.
All curves are normalized to match the observed $B-V$ color
(traditionally, UV extinction curves are plotted as
$E(\lambda-V)$/$E(B-V)$).  This illustrates the flatter nature of the
derived extinction curve (higher $R_V$) and weaker 2175~{\AA} bump
required along the GRB sightline relative to the average MW or
LMC sightlines.}
\label{fig:ext}
\end{figure}

We first attempted to fit using the general FM law (joined to the
standard CCM law in the rest-frame optical with a spline), leaving all
parameters free (except $c_1$ and $x_0$ as described above).
Unfortunately, because our observations do not extend far enough into
the rest-frame optical to properly constrain the optical/IR extinction
properties independent of the UV, the $R_V$ parameter is effectively
unconstrained in this case.  Fortunately, $R_V$ and $c_2$ also are
tightly correlated locally and can be tied together --- using, for
example, the correlation of \cite{Fitzpatrick1999} (linear) or that of
\cite{Reichart+2001} (quadratic, allowing for the optically flat,
steep-UV SMC-like curve).  Both correlations give acceptable (and very
similar) fits to our data, and the Fitzpatrick-constrained curve is
shown in Figures \ref{fig:sed}-\ref{fig:ext}.

We also attempted a range of non-FM models, such as those of
\cite{Calzetti+2000}, \cite{Maiolino+2004}, and \cite{Gaskell+2004}.
These curves all lack the 2175~{\AA} bump and do not fit the data.  In
addition, we tried to fit the multi-parametric extinction curve from
\cite{Li+2008}, which can incorporate the 2175~{\AA} bump and gives a
fairly reasonable fit (however, the $c_1$ parameter diverges and had
to be fixed manually, and the result is significantly worse than the
FM curve).  As the Li curve has not been used extensively on local
sightlines, it is difficult to interpret the results, and we will not
discuss it further.

The results from our various fits are presented in Table
\ref{tab:extfits}.  Note that despite the qualitative similarity of
the curve to MW and LMC sightlines, three major parameters ($R_V =
4.17 \pm 0.15$, $c_3 = 1.70 \pm 0.29$, and $c_4 = 0.28 \pm 0.07$)
differ significantly from the average values ($\gamma$ is consistent
with the average value).  In general, the GRB~080607 sightline is
UV-greyer, and its 2175~\AA\ bump weaker, than the average MW
sightline.  Still, all these properties are in the \textit{range} seen
along different sightlines locally (e.g., from \citealt{FM}: $2.3 <
R_V < 6.6$, $1.2 < c_3 < 4.5$, $0.15 < c_4 < 0.90$).  No single
local analog appears to match the properties seen toward the GRB exactly,
but it is nevertheless notable that our data are so well fit by the
standard, locally derived laws without the need for any unusual
parameters.\footnote{This is not simply a matter of the flexibility of
  the fitting function: the model is quite limited in scope, with only
  four free parameters, each of which is constrained to a small
  allowable range.  Indeed, some reported extragalactic sightlines,
  e.g., the high-$z$ QSO sightline of \cite{Maiolino+2004}, cannot be
  accurately fitted within this model.}  We will further discuss the
implications of the FM parameters in \S \ref{sec:environs}.

Assuming any particular extinction model is not strictly necessary for
this GRB: the large extinction column actually allows us to directly
{\it measure} the wavelength-dependent extinction without need for
fitting.  Traditionally, UV extinction curves are presented as
$E(\lambda-V)$ (i.e., $A_\lambda-A_V$; the optical extinction itself
$A_V$ need not be known).  Our $K_s$-band measurement corresponds to the
rest-frame $V$ band, and so if the intrinsic slope can be assumed, one
can simply measure this value for each filter (or wavelength bin) by
comparing the observed $\lambda$-$V$ color to the predicted color for
the assumed intrinsic spectrum.  The results are plotted in Figure
\ref{fig:ext}, illustrating the intrinsic differences between the
curves and the inability of most of them to fit the data.

\subsection{Effect of Varying Intrinsic $\beta$}
\label{sec:extbeta}

The above quoted results all assume $\beta = 0.7$.  In reality, we do
not know the exact intrinsic spectral index, which varies from burst
to burst. As previously mentioned, the extinction of this burst is
sufficiently large, and the intrinsic variation in $\beta$ between
events relatively small, that the errors introduced from variation in
the spectral index are small.  Here we quantify that statement and
propagate the effects into our parameter uncertainties.

\cite{Kann+2010} have compiled photometry for a large number of
bright, well-observed, {\it Swift}-era GRBs and performed fits to the
extinction (using the standard MW/LMC/SMC method) and spectral index
of each event.  We downloaded the data in Table 2 of that work and
removed all events which did not have a best-fit (among the three
models) $A_V < 0.2$ mag within 2$\sigma$ to exclude events with
significant or poorly determined extinction.  We further removed any
events reporting an unphysical $A_V < 0$ mag at more than $2\sigma$
and any event with an uncertainty in its derived spectral index
$\sigma_\beta > 0.2$.  The intrinsic spectral indices of this final
sample of 21 low-extinction, well-constrained bursts have an average
spectral index of $\beta = 0.70$ and standard deviation
$\sigma_{\beta} = 0.26$.  We take this as a representative sample with
which to determine a prior on the intrinsic (unextinguished) spectral
index $\beta$.

The observed spectral index between the $J$ and $K_s$ bands for this GRB
is $\beta = 3.5$, so the impact of reddening (between these
wavelengths) from dust is clearly much larger (by about an order of
magnitude) than the typical variation in the intrinsic spectral index.  This
variation in the intrinsic index is, however, the largest source of
uncertainty in the measurement of the extinction parameters.  To take
this into account, we refit our preferred extinction models for the
$\pm 1\sigma$ cases and combined the resulting variation of the
best-fit value in quadrature with the statistical uncertainties on the
$\beta = 0.7$ fit.  The final values for all extinction parameters
(using the Reichart $c_2 - R_V$ correlation; the Fitzpatrick
correlation is not significantly different) are presented in Table
\ref{tab:extpars}.

As an alternative to assuming an intrinsic optical $\beta$, we also
attempted our fits by including the X-ray flux value at the extraction
epoch and assuming an unbroken power law over the full range between
the optical and X-ray data (which allows for a much more precise
derivation of $\beta$ as well as a constraint on the overall flux
normalization, though it is strongly dependent on this assumption of
an unbroken intrinsic index).  This gives generally quite consistent
values with our optical-only fit, in further support of our assertion
that the derived dust properties are not strongly affected by our
assumptions about the intrinsic spectrum.

\begin{deluxetable*}{lllllllll}
\tablewidth{0pc}
\tablecaption{Extinction Fits}
\tablehead{ \colhead{Model} & \colhead{$A_V$} & \colhead{$R_V$} & \colhead{$c_1$} & \colhead{$c_2$} & \colhead{$c_3$} & \colhead{$c_4$} & \colhead{$\gamma$} & \colhead{$\chi^2/{\rm dof}$} \\
            \colhead{}      & \colhead{(mag)} & \colhead{}      & \colhead{}      & \colhead{}      & \colhead{}      & \colhead{}      & \colhead{}         & \colhead{} \\
}
\startdata
Average MW     &$1.25\pm0.03$& 3.1         & $-$0.07       &  0.70       & 3.23        & 0.41        & 0.99        &   127 / 24 \\
Average LMC    &$1.09\pm0.02$& 3.2         & $-$1.28       &  1.11       & 2.73        & 0.64        & 0.91        &   275 / 24 \\
LMC2           &$0.16\pm0.03$& 2.6         & $-$2.16       &  1.31       & 1.92        & 0.42        & 1.05        &  1143 / 24 \\
SMC            &$0 \pm 0.01$ & 2.73        & $-$4.96       &  2.26       & 0.37        & 0.46        & 0.99        &  1159 / 24 \\
CCM            &$0.82\pm0.06$&$2.41\pm0.12$&             &             &             &             &             &  123 / 22 \\
FM+tie         &$3.26\pm0.31$&$4.17\pm0.15$&$1.11\pm0.12$&$0.31\pm0.04$&$1.70\pm0.29$&$0.28\pm0.07$&$1.10\pm0.06$&  24.2 / 20 \\
FM+Reichart    &$3.52\pm0.35$&$4.69\pm0.19$&$1.29\pm0.15$&$0.30\pm0.05$&$1.66\pm0.30$&$0.31\pm0.07$&$1.07\pm0.07$&  22.9 / 20 \\
Li             &$1.70\pm0.06$&             &$200$       &$12.3\pm0.6$ &$14\pm285$&$0.03\pm0.01$ &             &  38.7 / 20 \\
\enddata
\tablecomments{Comparison of fits to the SED of GRB~080607 using a
  variety of extinction models, most of which cannot adequately fit
  the observations.  Because the optical spectrum and photometry
  dominate the observations, most models converge to a low extinction
  value to try to accommodate the weak 2175~{\AA} bump and seemingly
  flat spectrum.  These models are not consistent with the red IR
  color.  Both a high $R_V$ and a low $c_3$ are required to explain
  the optical and IR data together, as reflected in the FM fits.
  Parameter uncertainties do not include the effect of the uncertain
  intrinsic spectral index $\beta$ (a value of 0.7 is assumed).}
\label{tab:extfits}
\end{deluxetable*}

\begin{deluxetable}{lll}
\tablewidth{0pc}
\tablecaption{FM Extinction Parameters for GRB~080607}
\tablehead{ \colhead{}     & \colhead{Optical alone} &  \colhead{Optical + X-ray} \\
            \colhead{Parameter} & \colhead{value}     &  \colhead{value}}
\startdata
$\beta$  & 0.7  $\pm$ 0.26   & 1.08 $\pm$ 0.05 \\
\hline
$A_V$    & 3.26 $\pm$ 0.35   & 3.07 $\pm$ 0.32 \\  
$R_V$    & 4.17 $\pm$ 0.25   & 4.52 $\pm$ 0.23 \\  
$c_1$    & 1.11 $\pm$ 0.20   & 1.37 $\pm$ 0.15 \\  
$c_2$    & 0.31 $\pm$ 0.07   & 0.22 $\pm$ 0.05 \\  
$c_3$    & 1.70 $\pm$ 0.30   & 1.82 $\pm$ 0.32 \\  
$c_4$    & 0.28 $\pm$ 0.08   & 0.37 $\pm$ 0.08 \\  
$\gamma$ & 1.10 $\pm$ 0.07   & 1.07 $\pm$ 0.06 \\  
$x_0$    & 4.596  \\
\enddata
\tablecomments{Final FM extinction parameters for GRB~080607.  The
  values in the left column incorporate only the optical data and
  include the effect of unknown intrinsic spectral index.  Values at
  right assume an unbroken power law between the optical and X-rays.  
  $c_1$ and $c_2$ are tied to $R_V$ as described in the text; $R_V$ is
  significantly higher than the average MW or LMC curve but has
  a typical value for dense sightlines.  The 2175~{\AA} bump (strength
  given by $c_3$), ubiquitous in the MW but nearly absent in
  the SMC, is present but weaker than in the MW or LMC.  }
\label{tab:extpars}
\end{deluxetable} 

\subsection{X-ray Scattering?}
\label{sec:xscat}

Of particular note in Table \ref{tab:extpars}, and consistent with our
previous work \citep{Prochaska+2009}, is the conclusion of a large
extinction column ($A_V = 3.26 \pm 0.35$ mag).  This identification of
GRB~080607 as a highly extinguished event makes it a potentially
useful test case of the X-ray scattering model for early-time
afterglows \citep{Shen+2009}.  However, even $A_V = 3$ mag is
generally inadequate to expect any significant effects on the X-ray
light curve in this case.  Following the discussion by
\cite{Shen+2009}, we calculate the 1~keV specific fluence from the
prompt emission using the parameters given in \cite{GCN7862} and
integrate the X-ray afterglow flux (starting at 100~s, and ignoring
the X-ray flare) using our power-law fit.  The resulting ratio of
$S_{\rm AG}/S_{\rm prompt}$ = $(1.0 \times 10^{-6}) / (1.6 \times
10^{-7}) = 6.3$ places an upper limit on the scattering opacity at
this wavelength (i.e., $\tau_{\rm scat} < 6.3$).  Translating this to
a limit on the optical opacity using Equation (8) of \cite{Shen+2009},
the limiting dust extinction for this case is the thoroughly
unconstraining $A_V < 686$ mag, a value about 200 times higher than
our direct measurement.  Equivalently, the total fluence of the
dust-scattered X-rays for this event is anticipated to be 200 times
lower than the actual afterglow fluence observed, and therefore
undetectable.  Indeed, for this event, the X-ray light curve follows a
simple unbroken power law and (with the exception of the early X-ray
flare, a prompt-emission feature;
\citealt{Kocevski+2007,Chincarini+2007}) no significant hardness
variations of the type predicted by \cite{Shen+2009}.  We conclude
that, despite the large surrounding dust column, X-ray scattering is
not significant for this GRB.

\section{Discussion}

\subsection{Afterglow Luminosity in Context}
\label{sec:lum}

The impressive optical brightness for an event at $z=3$ has already
been noted.  In fact, as we shall show, after the effects of
extinction are taken into account, GRB~080607 is among the most
optically luminous GRBs (and therefore objects of any sort) to date,
second only to GRB~080319B.

Following \cite{Kann+2007,Kann+2010}, we select as our comparison
filter the $z=1$ $R$ band (that is, the wavelength which is shifted to
the observed $R$ band if at $z=1$); this corresponds roughly to the
rest-frame $U$ band.  For GRB\ 080607, this is shifted all the way to
approximately the observed $J$ band.  Therefore, taking advantage of
the apparent lack of color change, we shift all other filters to the
$J$-band light curve using our model fluxes, extending this curve back
to the observed emission peak.  This curve is corrected for Galactic
extinction (only 0.02 mag), for host extinction (4.94 mag), and for
the difference in luminosity distance between $z=1$ and $z=3.036$
using standard cosmology ($h$=0.7, $\Omega_M = 0.3$,
$\Omega_\Lambda=0.7$).  A small K-correction is then applied to match
the spectrum exactly with the $z=1$ $R$ band, and the light curve is
scaled (undilated) to $z=1$.

\begin{figure}
\centerline{
\includegraphics[scale=0.7,angle=0]{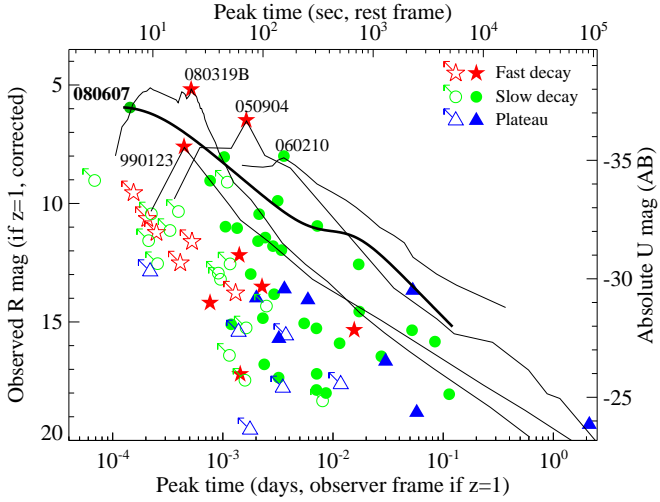}}
\caption{Near-UV luminosity of GRB~080607 (bold curve) compared to
  several other prominent bursts, as well as to a large sample of
  rapidly observed {\it Swift} GRBs from \cite{Kann+2010}.  Colored points
  indicate peak observed luminosities of the events described in that
  paper (unfilled points are events caught after the peak and
  therefore only lower limits on the peak luminosity.)  At peak, GRB
  080607 is among the most luminous GRBs known, peaking at $M_U \approx
  -37$ mag.  At $z=1$ it would peak at mag 6 if unobscured by dust.}
\label{fig:lum}
\end{figure}

The result is plotted in Figure \ref{fig:lum}, compared with the light
curves of GRB~080319B and the three next most luminous events (from
Figure 7 of \citealt{Bloom+2009}), and with the peak luminosities of a
large sample of well-studied {\it Swift} bursts (from Figure 7 of
\citealt{Kann+2010}.)  At the beginning of observations GRB~080607 is
comparable in luminosity to GRB~080319B (and at early times was likely
brighter), but the prompt optical flaring of GRB~080319B pushes that
burst to a higher luminosity over the next several minutes.  GRB~080607 
remains among the five most luminous bursts for the rest of its
observed evolution.  This illustrates the remarkable attributes of
this burst that allowed it to provide such a detailed analysis of its
environment.  In terms of the afterglow (external shock) emission
alone, GRB~080607 may yet be the most luminous: GRB~080319B's peak
appears to correlate with its prompt emission and fades particularly
rapidly when the prompt emission ends; the origin of its early-time
optical emission is still debated \citep{Racusin+2008}.  The optical
light curve of GRB~080607 bears no relation to the prompt emission and
is certainly external shock-dominated at all times.

\begin{figure}
\centerline{
\includegraphics[scale=0.7,angle=0]{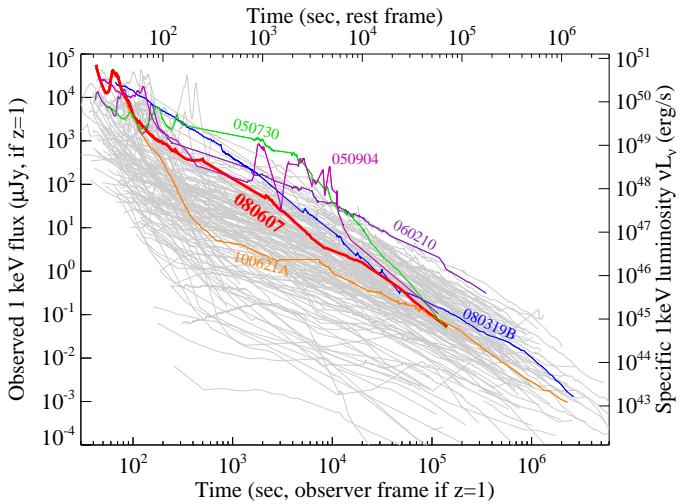}}
\caption{X-ray luminosity of GRB~080607 (bold red curve) compared to
  all other {\it Swift}-followed GRBs.  Several other prominent bursts
  are also individually colored.  GRB~080607 is among the most X-ray
  luminous bursts at peak, but fades quickly to an average luminosity
  by later times.}
\label{fig:xlum}
\end{figure}

\subsection{Physical Properties}
\label{sec:physics}

Unfortunately, the physical properties responsible for making
GRB\ 080607 so energetic remain mostly hidden from view.  In the burst
rest frame, our optical observations extend only to $10^3$~s and the
X-ray observations cease at $t \approx 1$ day, which does not usefully
constrain the jet opening angle.  Conservatively setting $t_{\rm jet}
> 6 \times 10^4$~s using the X-ray light curve, following the standard
equations for the jetting time \citep{Sari+1999,Frail+2001} and
fiducial values of density $n = 100$ cm$^{-3}$ \footnote{This is an
  unusually large value of $n$, motivated by the apparent low value of
  the cooling break $\nu_c$ and inference of a dense molecular
  environment along the line of sight, as discussed later in this section.  
  Fortunately, the value of
  $n$ only weakly affects the derived value of $\theta$ and
  $E_{\gamma}$.} and efficiency $\eta = 0.2$, we measure a jet opening
angle of $\theta_{\rm jet} > 3.6 \eta_{0.2}^{1/8} n_{100}^{1/8}$
degrees; the equivalent lower limit on the beaming-corrected gamma-ray
energy release is $E_{\gamma} > 1.8 \times 10^{51} \eta_{0.2}^{1/4}
n_{100}^{1/4}$ erg, a fairly typical value.  It is therefore not clear
whether the extreme apparent luminosity of this burst is attributable
to intrinsically large energetics \citep{Cenko+2010a}, favorable
viewing angle (of a nonuniform, centrally concentrated jet, as was
suggested for GRB~080319B by \citealt{Racusin+2008}), an intrinsically
narrowly concentrated (uniform) jet, or some combination of these
parameters.

To a large extent, the optical luminosity is simply another reflection
of the total energetics of the burst itself: both in theory
\citep{Sari+1998} and observationally
\citep{Gehrels+2008,Nysewander+2009,Kann+2010}, the inferred afterglow
luminosity scales approximately linearly with $E_{\rm iso}$, and if
the optical light curve is extrapolated to late times the predicted
optical flux is in the middle of the fluence-normalized distribution.
However, there is more to the story: the X-ray light curve of this
burst is (except at the earliest times) not particularly bright; 
when normalized to the burst fluence
it is quite typical for a {\it Swift} burst at early times (and
actually is unusually faint at late times, due to its rapid unbroken
decay).

There are two broad ways to interpret this.  The simplest
interpretation is that the cooling-break frequency $\nu_c$ has a
particularly low value compared to most GRBs, perhaps even below the
optical band (the available data are marginally consistent with the
X-ray and optical bands being on a single spectral power law).  The
obvious culprit for this involves the external density $n$: The X-ray
flux should be independent of density (assuming that $\nu_{\rm X} >
\nu_c$ for most bursts), but the cooling break and optical flux are
sensitive to it ($\nu_c \propto n^{-1}$; below the cooling break $F
\propto n^{1/2}$).  An external density 10 or 100 times the
``typical'' {\it Swift} value would push the cooling break from its
typical position between the optical and X-ray bands into or below the
optical band at early times, increasing the optical luminosity.
Indeed, after correcting for extinction the early-time broadband SED
appears to demand a low-cooling break: the optical-to-X-ray index at
only 300~s is $\beta_{OX} = 1.1$, consistent with the X-ray spectral
slope ($\beta_{\rm X} = 1.16 \pm 0.13$).  The probable low value of the cooling break also helps explain
why a similar extinction column is derived whether the optical data
is considered alone (the most general case) or in conjunction with 
the X-ray data assuming an unbroken power-law (which requires $\nu_c < \nu_{\rm opt}$),
as demonstrated in \S \ref{sec:extbeta}.  Unfortunately, the period of
simultaneous temporal coverage between the optical and X-ray observations
is too short to determine, via the light curve, whether a break is present 
between the bands.
(The burst could also have exploded into a wind-stratified medium ---
one with variable density $n \propto r^{-2}$ --- in which case $\nu_c$
rises with time and the optical flux fades more rapidly than the X-ray
flux, as is observed.)

The alternative interpretation is that the optical flux originates
from a separate emission component, the most obvious candidate being
the reverse shock \citep{Meszaros+1997}.  Several previous early
fast-fading light curves have been associated with reverse shocks
\citepeg{Akerlof+1999,Kobayashi+2003,Perley+2008,Steele+2009};
qualitatively, the behavior of GRB~080607 appears similar to these
events, although the decay is somewhat slower and there are no
late-time observations to determine whether the light curve became
forward-shock dominated as predicted.  The factors determining the
luminosity of the reverse shock \citep{Zhang+2003} are generally the
same as for the forward shock (and so a high external density would
similarly aid the production of a luminous afterglow), but can be
further amplified if the magnetization $R_B =
\epsilon_{B,r}/\epsilon_{B,f}$ of the reverse shock is high, due (for
instance) to primordial fields in the ejecta \citep{Gomboc+2008}.

Unfortunately, the lack of late-time observations (to search for the
appearance of a forward shock) or radio data (to more directly
constrain $n$) prevents us from distinguishing between these
possibilities.  Fortunately, we can speak more confidently about the
other aspect of this burst's remarkable luminosity: the fact that it
peaked so early (even if two bursts have similar energetics and
late-time luminosities, the power-law nature of GRB light curves
ensures that the event with the earlier peak time will have
significantly larger peak brightness, fleeting as it is.)

The peak time (for $\nu > \nu_m$) is set by the deceleration timescale
of the ejecta \citep{Sari+1999}. Because the afterglow has already
peaked and is fading at the start of our observations, the ejecta must
have accumulated enough circumstellar matter to begin to decelerate
and develop an external shock by this time: a mere 24~s after the BAT
trigger (6~s in the rest frame), or more conservatively 32~s after the
beginning of the prompt emission (8~s in the rest frame).

Such rapid deceleration generally requires a high initial Lorentz
factor $\Gamma$, although a very high interstellar density also contributes.
Using Equation 3 of \cite{Rykoff+2009}\footnote{This equation is
  strictly valid only for the thin-shell scenario, in which the burst
  duration is less than the deceleration time
  \citep{Sari+1999,Meszaros+2006}.  This is not strictly true for this
  GRB, as prompt emission is observed to continue during the
  light-curve decline.  However, as noted previously, the energetics
  are dominated by a single, bright pulse which ends well before the
  start of optical observations.}, we estimate $\Gamma > 660
\eta_{0.2}^{1/8}n_{100}^{-1/8}$, where $\eta_{0.2}$ and $n_{100}$
indicate values of the efficiency and external density relative to
fiducial values of 0.2 and 100 cm$^{-3}$, respectively (see also
\citealt{Molinari+2007}).  Based on the preceding discussion of the
late-time optical luminosity, we have chosen an unusually large value
for the interstellar density; even in this case the constraint on $\Gamma$
is at the top end of the afterglow-inferred range (if still somewhat
below the pair-opacity limits recently provided by the 
{\it Fermi}-LAT: \citealt{Abdo+2009b}).  It is notable that both
$E_{\rm iso}$ and $\Gamma$ are exceptionally large for this burst,
which could suggest that the properties may be correlated.

\subsection{X-Ray and Optical Properties: the Environment of GRB 080607}
\label{sec:environs}

The derivation of precise values for the extinction parameters along
the GRB~080607 sightline (Table \ref{tab:extfits}) gives us an
additional means for learning about its host environment.  Although the
reason for the variation of these parameters is not well understood
even within the MW, some broad conclusions can be drawn.

First, we note the high value of $R_V \approx 4$ (or equivalently,
since the parameters are tied in our modeling, the small value of
$c_2$), indicating a relatively flat extinction curve.  In the diffuse
interstellar medium (in the MW and in other galaxies as well),
$R_V$ typically takes on lower values of 2--4.  UV-flat extinction
curves are generally restricted to denser sightlines, probably because
grains are able to coagulate to larger sizes \citep{Valencic+2004}.
(At the same time, however, dense regions can also have low values of
$R_V$ as well as high ones.)  The high $R_V$ value is therefore
suggestive of a dense environment --- fully consistent with the
conclusion from the atomic and molecular analysis that the sightline
penetrates through a dark molecular cloud in its host.

Second, the value of $c_3$ is nonzero, indicating a significant
2175~\AA\ absorption bump.  This is one of only a few clear detections
of this feature at cosmological distances
\citep{Motta+2002,Junkkarinen+2004,Ellison+2006,Kruhler+2008,Eliasdottir+2009}
and the highest-redshift detection of the feature yet.  The identity
of the carrier is still unknown (although polycyclic aromatic
hydrocarbons and graphite are considered promising candidates; see
\citealt{Draine2003} for a review) and the processes that cause it to
be present or absent are similarly not yet certain: an evolved stellar
population \citep{Noll+2007}, metallicity \citep{Fitzpatrick2004}, the
strength of the UV radiation field \citepeg{Gordon+1997}, and
disturbance of the environment due to shocks \citep{Seab+1983} have
all been cited in explaining its absence.  Generically, however, it
seems to be present in almost all sightlines in the MW and LMC,
and in nearby disk galaxies, but absent in more disturbed locations
such as the SMC, nearby starburst galaxies \citep{Gordon+2005}, and at
least one highly disturbed sightline within the MW
\citep{Valencic+2003}.  This suggests that the interstellar medium of
the host of GRB~080607 is a closer analog of the more quiescent
environments found in the MW and LMC than of the extreme
conditions of nearby galaxies having high specific star-formation
rates. Our upcoming study of the host galaxy (Chen et al. 2011) may
help test this hypothesis.

\begin{figure*}
\centerline{
\includegraphics[scale=0.8,angle=0]{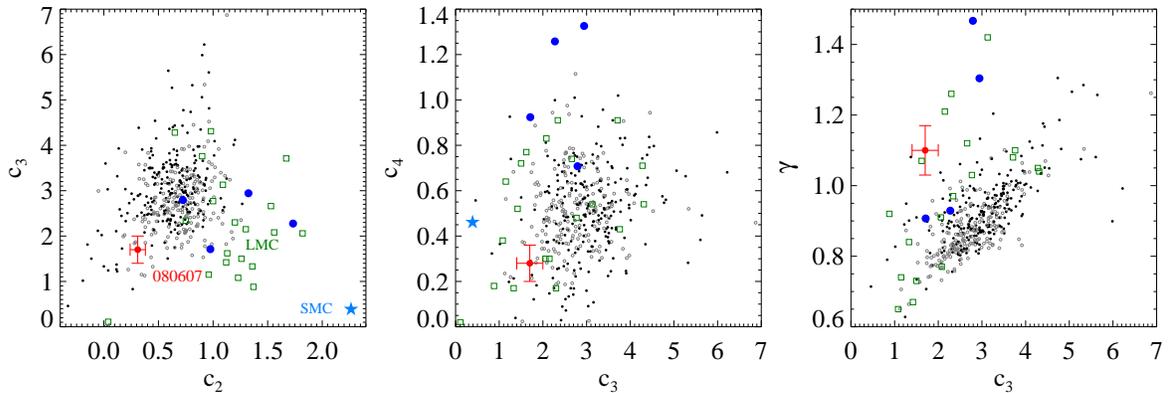}}
\caption{Extinction parameters for GRB\ 080607, compared to various
  MW and LMC sightlines from \cite{Valencic+2004} and
  \cite{Misselt+1999}.  Diffuse MW sightlines are indicated
  with small, gray open circles; dense sightlines are indicated with
  small, filled black circles.  Peculiar MW sightlines
  incompatible with the standard CCM one-parameter family are
  identified as blue circles.  LMC sightlines are indicated with
  rectangles; the SMC curve is plotted as a star.  Extinction
  parameter $c_2$ is a measure of the UV slope (inverse greyness);
  $c_3$ is a measure of the strength of the 2175~{\AA} bump and
  $\gamma$ is a measure of its width; and $c_4$ indicates the strength of
  the far-UV rise.  All parameter values (and all pairs of two values)
  are within the distribution seen locally, although there is no
  single example of a local sightline that is consistent with the
  extinction properties of the GRB~080607 sightline in all aspects.}
\label{fig:extpar}
\end{figure*}

The strength of the bump is, however, weaker than in almost any
sightline in either the MW or LMC (Figure \ref{fig:extpar}).
Furthermore, the degree to which the bump is weaker does \emph{not}
follow the local correlations: in the MW, very low values of $c_3$
tend to correlate with very low values of $c_2$ (weakly) and $\gamma$
(strongly).  In our case, a low $c_2$ is observed, but it is still
much higher than for the Orion Nebula sightlines in which the lowest
values of $c_3$ are seen.
This may be an indicator that a different phenomenon is suppressing
this carrier than is in operation within the MW Galaxy.  Metallicity 
is not likely the culprit: the molecular cloud giving rise to the
observed extinction has near-Solar metallicity despite being at $z >
3$ \citep{Prochaska+2009}.

The strength of the FUV rise, $c_4$, is fairly typical for local
sightlines.  However, the origin of the rise is even less secure than
that of the 2175~{\AA} bump, and does not significantly constrain the
environment.

The X-ray inferred host-galaxy equivalent hydrogen column of GRB
080607 was measured to be $N_{\rm H} = 2.7^{+0.8}_{-0.7} \times
10^{22}$ cm$^{-2}$, and comparable to the neutral hydrogen column
$N_{HI} = 1.5^{+0.6}_{-0.5}\times10^{22}$ derived from the damped
Lyman-$\alpha$ line \citep{Prochaska+2009}.  This is a very large
value, even considering the high extinction in this direction: the
ratio of $N_{\rm H}/A_V = 8 \times 10^{21}$ cm$^{-2}$/mag is several times larger
than observed in the MW, although quite typical of GRBs for which both
values have been securely measured \citepeg{Schady+2010}.  It is
possible that this arises for reasons unrelated to the molecular cloud
--- for example, if additional dust-free gas is located closer or
further along the sightline relative to the molecular cloud that is
responsible for the absorption.  If intrinsic, this combination of a
weak (but present) bump and a large $N_H/A_V$ ratio is consistent with
the correlation discussed by \cite{Eliasdottir+2009} and
\cite{Gordon+2003}.

\section{Conclusions}
\label{sec:conclusions}

One of the brightest and best-studied GRBs (at early times) of the
{\it Swift} era, GRB~080607 holds particular potential for revealing
the nature of GRBs and their environments at high redshift.  While the
relatively limited observed temporal range restricts our ability to
study the intrinsic nature of this event, this is more than
compensated by the abundant early-time optical/IR data that reveal the
detailed properties of the dark-cloud sightline in its distant host.

The utility of this event is perhaps most evident in the context of
the class of ``dark'' GRBs.  Many factors, both intrinsic (high
$E_{\rm iso}$, $\Gamma$, and $n$) and extrinsic (large but not
extremely large $A_V$, a redshift placing the 2175~{\AA} bump in the
optical window, and the fortuitous ability to observe immediately with
telescopes in both the continental US and Hawaii) had to conspire
together to allow an event to be observable in such rich detail.  Had
this event been slightly less luminous (``only'' comparable to GRB
990123, $\sim$2 mag fainter at most epochs), its afterglow would have
been only marginally detected, and only at the earliest times;
further decrease in luminosity would have rendered it undetectable to
small telescopes.  Even a
modest increase in the amount of extinction (higher by $A_V \approx
1$--2 mag) or the presence of relatively UV-opaque SMC-like dust would
have a similar impact, suppressing all of the optical measurements.

The literature contains many examples of such sources: GRBs with a
comparable dust column but insufficient luminosity to shine through it.
Some prominent cases include GRBs 970828 \citep{Djorgovski+2001}, 060923A
\citep{GCN8698}, 061222A and 070521 \citep{Cenko+2009,Perley+2009},
070306 \citep{Jaunsen+2008}, 081221 \citep{Tanvir+2008}, and 090709A
\citep{Cenko+2010b}.  But even these objects were unusually bright or
had particularly rapid or deep observations in their favor.  A truly
typical-luminosity {\it Swift} event without rapid or deep
observations would completely escape notice in most cases, permitting
only shallow limits on its extinction column.  Therefore, there is
every reason to think that very dusty environments like that of GRB
080607 are actually not uncommon among GRBs (if not necessarily
ubiquitous).  This is in agreement with our afterglow plus host
survey with the P60 and Keck telescopes \citep{Cenko+2009,Perley+2009}.

The extinction curve along the GRB~080607 sightline --- a dark
molecular cloud at $z=3.036$ --- is quite similar to that of our own
Galaxy (with a significant 2175~{\AA} bump), though there are
differences in finer details.  The success in modeling the extinction
curve of this event within entirely locally developed models is in
some ways heartening, giving us confidence that with sufficient
knowledge we should be able to understand the absorption properties
even out to these immense distances.  As perhaps the most detailed
determination of the extinction properties of a galaxy at cosmological
redshift to date, we suggest that the extinction curve in this work
may be of use to others attempting to take into account the effects in
other galaxies at high redshift (see Appendix).  At the same time,
some other GRBs and other techniques have also at times pointed to
extinction curves that diverge dramatically from local templates, so
the topic should continue to be addressed with caution.

Once corrected for extinction, GRB~080607 rivals the ``naked eye
burst'' GRB~080319B as the most luminous known object in the Universe.
This extreme early luminosity of GRB~080607 is likely the product of a
variety of factors: it has one of the largest $E_{\rm iso}$ values to
date, and its optical luminosity may have been further amplified by a
large circumburst density in its host (or, alternatively, a bright
reverse shock).  Unlike GRB~080319B, GRB~080607 has a smooth optical
peak and shows no correlation with prompt emission at that time.
Events like GRB~080607 demonstrate the power of GRBs to illuminate the
darkest corners of the Universe: not just the reionization era (on
which much current attention is focused) but also the dustiest regions
over the following several billion years when the global star formation rate
--- much of it occurring behind optically thick dust clouds --- was at
its maximum.  Such dust-obscured regions are extremely difficult to
study by other techniques, or even with most GRBs, as demonstrated by
the class of ``dark'' bursts.  The combination of early observations
and extreme energetics of GRB~080607 were enough to overcome even this
difficulty, and demonstrate the power of rare, individual events to
illuminate these hard-to-study regions and improve our understanding
of the early universe.

\acknowledgments


J.S.B.'s group is partially supported by a grant from the NSF (award
0941742).  The Gamma-Ray Afterglows As Probes (GRAASP) collaboration
is supported by NASA/{\it Swift} Guest Investigator grants NNX08AN90G
and NNX09AO99G.  A.V.F., S.B.C., and W.L. acknowledge generous financial
assistance from Gary \& Cynthia Bengier, the Richard \& Rhoda Goldman
Fund, NASA/{\it Swift} grants NNX09AL08G and NNX10AI21G, the TABASGO
Foundation, and NSF grant AST-0908886.  A.N.M. would like to
acknowledge support from a NSF Graduate Research Fellowship.

PAIRITEL is operated by the Smithsonian Astrophysical Observatory
(SAO) and was made possible by a grant from the Harvard University
Milton Fund, a camera loan from the University of Virginia, and
continued support of the SAO and UC Berkeley. The PAIRITEL project is
further supported by NASA/{\it Swift} Guest Investigator grant
NNX08AN84G.

Some of the data presented here were obtained at the W. M. Keck
Observatory, which is operated as a scientific partnership among the
California Institute of Technology, the University of California, and
NASA.  The Observatory was made possible by the generous financial
support of the W. M. Keck Foundation.  We wish to extend special
thanks to those of Hawaiian ancestry on whose sacred mountain we are
privileged to be guests.

KAIT and its ongoing operation were made possible by donations from
Sun Microsystems, Inc., the Hewlett-Packard Company, AutoScope
Corporation, Lick Observatory, the NSF, the University of California,
the Sylvia \& Jim Katzman Foundation, and the TABASGO Foundation.

This work made use of data supplied by the UK Swift Science Data 
Centre at the University of Leicester.  This research also made 
use of the NASA/IPAC Extragalactic Database (NED) which is operated 
by the Jet Propulsion Laboratory, California Institute of Technology, 
under contract with the National Aeronautics and Space Administration. 

We thank S.~Klose for additional computations that confirmed our
luminosity results, and for the $H_2^*$ model developed by B.~T.~Draine
and supplied by Y.~Sheffer.  We are grateful to the staffs of the
observatories at which we obtained observations.  Finally, we wish to
acknowledge the hard work and dedication of the \Swift team, whose
successful mission has made this study possible.

\bigskip

{\it Facilities:} \facility{Swift}, \facility{Keck:I (LRIS)},
\facility{PAIRITEL}, \facility{SuperLOTIS}, \facility{ROTSE},
\facility{Lick:KAIT}, \facility{PO:1.5m}

\appendix

\section{Applying the GRB 080607 Extinction Law}

Because the extinction law we derive for GRB~080607 is contained
within the broader Fitzpatrick parameterization, application of this
extinction law is relatively straightforward.  Using the function {\tt
  fm\_unred.pro} in the GSFC IDL library, the extinction law can be
calculated using, for example, the following simple code.

\begin{verbatim}
function extinction080607, wav, rv, av

  ; Returns the extinction (in magnitudes) as a function of rest wavelength 
  ; (in Angstroms) for the GRB080607 sightline, normalized to A_V.

  if n_elements(rv) eq 0 then rv = 4.17
  if n_elements(av) eq 0 then av = 1.0
  ebv = av/rv
  influx = replicate(1., n_elements(wav))

  fm_unred, wav, influx, -ebv, extflux, r_v=rv, c3=1.70, c4=0.28, gamma=1.10
  return, -2.5*alog10(extflux)

end
\end{verbatim}

\noindent
Here $R_V$ is a free parameter, left to vary in principle --- although
use of the default value is strongly recommended, since in this
implementation it is tied to the $c_1$ and $c_2$ parameters within
{\tt fm\_unred.pro}.

\end{document}